# Title: Oxygen fugacities of extrasolar rocks: evidence for an Earth-like geochemistry of exoplanets

**Authors:** Alexandra E. Doyle[1]*, Edward D. Young[1], Beth Klein[2], Ben Zuckerman[2], Hilke E. Schlichting[1,2,3].

**Affiliations:**

[1]Department of Earth, Planetary, and Space Sciences, University of California, Los Angeles, Los Angeles, CA, USA.

[2]Department of Physics & Astronomy, University of California, Los Angeles, Los Angeles, CA, USA.

[3]Department of Earth, Atmospheric and Planetary Sciences, Massachusetts Institute of Technology, Cambridge, MA, USA.

*Correspondence to: a.doyle@ucla.edu or eyoung@epss.ucla.edu.

**Abstract:** Oxygen fugacity is a measure of rock oxidation that influences planetary structure and evolution. Most rocky bodies in the Solar System formed at oxygen fugacities approximately five orders of magnitude higher than a hydrogen-rich gas of solar composition. It is unclear whether this oxidation of rocks in the Solar System is typical among other planetary systems. We exploit the elemental abundances observed in six white dwarfs polluted by the accretion of rocky bodies to determine the fraction of oxidized iron in those extrasolar rocky bodies and therefore their oxygen fugacities. The results are consistent with the oxygen fugacities of Earth, Mars, and typical asteroids in the Solar System, suggesting that at least some rocky exoplanets are geophysically and geochemically similar to Earth.

**One Sentence Summary:** Elemental abundances from white dwarfs are used to calculate the Earth-like oxidation states of extrasolar rocks accreted by the stars.

**Main Text:** Estimating the composition of extrasolar planets from host-star abundances or from planet mass-radius relationships is difficult and unreliable (*1, 2*). The elemental abundances in some white dwarfs (WDs) provide an alternative, more direct approach for determining the composition of extrasolar rocks. White dwarfs are the remnant cores left behind when a star ejects its hydrogen-rich outer layers following the red giant phase. These remnant cores are ~ 0.5 solar masses ($M_\odot$) and about the same radius as Earth, are no longer powered by fusion, and slowly cool over time. Because of their high densities, and thus strong gravitational fields, elements heavier than helium rapidly sink below their surfaces, becoming unobservable. Nonetheless, spectroscopic studies show that up to half of WDs with effective temperatures

<25,000 K are "polluted" by elements heavier than He in their atmospheres (*3-5*). The source of these heavy elements is exogenous, coming from accretion of debris from rocky bodies that previously orbited the WDs (*6-9*). We exploit this pollution to measure the elemental constituents of extrasolar rocky bodies. We collated observations from the literature of polluting elements in six white dwarfs: SDSS J104341.53+085558.2 (*10*), SDSS J122859.92+104033.0 (*9*), SBSS 1536+520 (*11*), GD 40 (*8, 12*), SDSS J073842.56+183509.6 (*13*) and LBQS 1145+0145 (*14*) (hereafter SDSS J1043+0855, WD 1226+110, WD 1536+520, GD 40, SDSS J0738+1835 and WD 1145+017). Coordinates can be found in Table S1. Bulk compositions of the bodies polluting WDs resemble those of rocky bodies in the Solar System (*15, 16*) (Fig. 1).

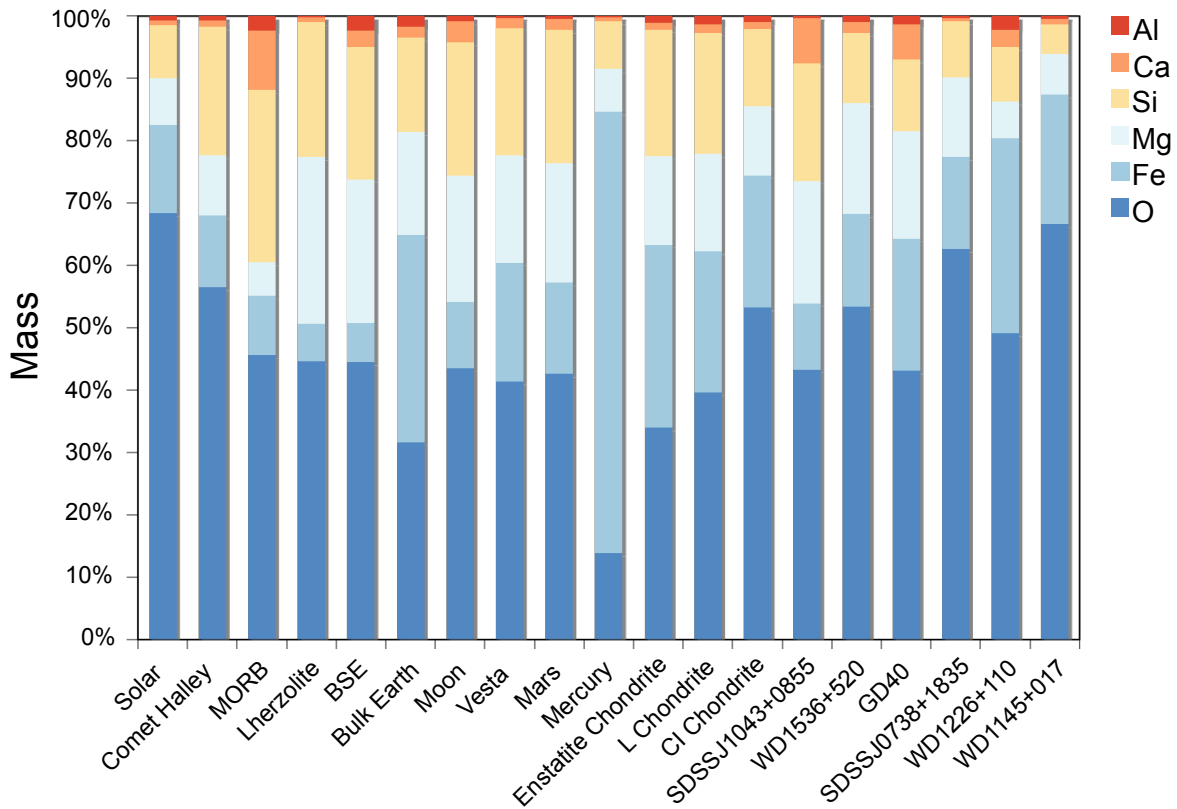

**Fig. 1. Bulk compositions by mass for six white dwarfs compared to Solar System bodies.** Bulk compositions of the six rock-forming elements Al, Ca, Si, Mg,

Fe and O are indicated by the colored bars. The six white dwarfs are shown in the right-most columns. Shown for comparison are Solar System objects: the Sun, Comet Halley (1P/Halley), Earth, the Moon, Vesta, Mars, Mercury, three types of meteorites (enstatite chondrite, ordinary L chondrite, and carbonaceous CI chondrite), and three terrestrial igneous rock types: mid ocean ridge basalt (MORB), lherzolite (representing Earth's mantle) and bulk silicate Earth (BSE) (*16*). The relatively high abundances of Fe in bulk Mercury and bulk Earth are due to their metal cores. The compositions of the white dwarfs are similar to the Solar System rocks. The large amount of O in WD 1145+017 is highly uncertain. Values are listed in Data S1.

We use the relative abundances of rock-forming elements in polluted WDs to determine the effective partial pressure of oxygen, i.e. the oxygen fugacity ($f_{O_2}$) of the accreted rocks. Oxygen fugacity is a measure of the degree of oxidation in the rocks. It corresponds to the effective partial pressure of gaseous oxygen that would be in thermodynamic equilibrium with the material of interest. In combination with other factors, the intrinsic oxygen fugacity of a planet will determine the relative size of its metallic core, the geochemistry of its mantle and crust, the composition of its atmosphere, and the forces responsible for mountain building (*17, 18*). Oxygen fugacity is also thought to be among the parameters that determine the habitability of a planet (*19*). In practice, $f_{O_2}$ is usually expressed as the non-ideal partial pressure of oxygen relative to a convenient reference value.

Oxygen fugacities of rocky planets are often reported relative to the reference Iron-Wüstite (IW) equilibrium reaction Fe (Iron) + ½ $O_2$ = FeO (Wüstite), such that $\Delta IW \equiv \log(f_{O_2}) - \log(f_{O_2})_{IW}$ (*16*). When expressed this way, differences in oxygen fugacity are nearly independent of temperature and pressure (*16*). The initial oxidation state of a rocky body with at least some Fe metal at the time of its formation is recorded by the concentration of oxidized iron (which we hereafter denote as FeO, although it may include other oxides of iron) in the rock and the concentration of Fe in the metal:

$$\Delta \text{IW} = 2\log\left(\frac{x^{\text{rock}}_{\text{FeO}}}{x^{\text{metal}}_{\text{Fe}}}\right) + 2\log\left(\frac{\gamma^{\text{rock}}_{\text{FeO}}}{\gamma^{\text{metal}}_{\text{Fe}}}\right), \tag{1}$$

where $x_i^k$ are mole fractions of the species $i$ in phase $k$, $\gamma_i^k$ are activity coefficients for the species, and thermodynamic activities are $a_i^k = x_i^k\,\gamma_i^k$. To facilitate comparison, we set the uncertain activity coefficients to unity, so the second term on the right-hand side of Equation (1) vanishes. Equation (1) expresses the $f_{\text{O}_2}$ at the time the planet or planetesimal formed (*20, 21*); we refer to this as the "intrinsic" oxygen fugacity of the body. The partitioning of iron between rock and metal during formation leaves a record of the intrinsic oxygen fugacity in the form of the mole fraction of FeO in the rocks, $x^{\text{rock}}_{\text{FeO}}$. This signature persists even after the rock and metal are separated by the process of differentiation (partitioning between core and mantle). This is because changes in the valence state of iron during subsequent reactions proceed without appreciably altering the total amount of iron bonded to oxygen in the rocks. For example, the reaction

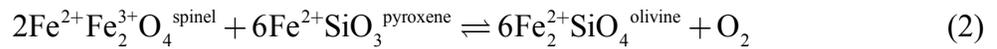

$$2\text{Fe}^{2+}\text{Fe}^{3+}_2\text{O}_4{}^{\text{spinel}} + 6\text{Fe}^{2+}\text{SiO}_3{}^{\text{pyroxene}} \rightleftharpoons 6\text{Fe}^{2+}_2\text{SiO}_4{}^{\text{olivine}} + \text{O}_2 \tag{2}$$

determines the $\text{Fe}^{3+}/\text{Fe}^{2+}$ ratio, and thus the $f_{\text{O}_2}$, in a rock containing the minerals spinel, pyroxene and olivine, without substantially altering the total Fe bonded to oxygen (from 2.17 oxygens per Fe to 2.00 oxygens per Fe). Reactions like these after the formation of a rocky body lead to local variations in $f_{\text{O}_2}$ within the body but do not generally alter the intrinsic oxygen fugacity recorded by application of Equation (1) (Fig. 2). The intrinsic oxygen fugacity of Earth is constrained by $x_{\text{FeO}}^{\text{mantle}} = 0.06$ (8 weight percent FeO) in its mantle and the composition of its Fe-rich core. This leads to a terrestrial $\Delta$IW value of about $-1$ to $-2$, with the range due to the uncertain values for the activity coefficient ratio [commonly used values of $\gamma_{\text{FeO}}^{\text{mantle}} / \gamma_{\text{Fe}}^{\text{core}}$ range from $\sim 1$ to 4 (*22*)].

The material accreted by the six polluted white dwarfs in this study are rocks devoid of metal, as demonstrated by the lack of excess Fe relative to oxygen (Fig. 1). Separation of metal and rock during accretion onto WDs is suggested by the ranges in element ratios in polluted WDs (*23*) and from observations of a metal-density planetesimal core orbiting a WD (*24*). Our $f_{\text{O}_2}$ measurements are representative of the fugacity values at the time of core formation, even though they are derived from crustal or mantle rocks. The maximum intrinsic oxygen fugacity calculated from Equation (1) approaches 0 as the mole fraction of Fe in the silicate increases. Values for $\Delta$IW greater than $\sim -0.9$ for elemental concentrations similar to Solar System rocks imply that all of the iron has been oxidized, and that the intrinsic $\Delta$IW from Equation (1) are therefore minimum estimates for the oxidation state at the time the rocks formed.

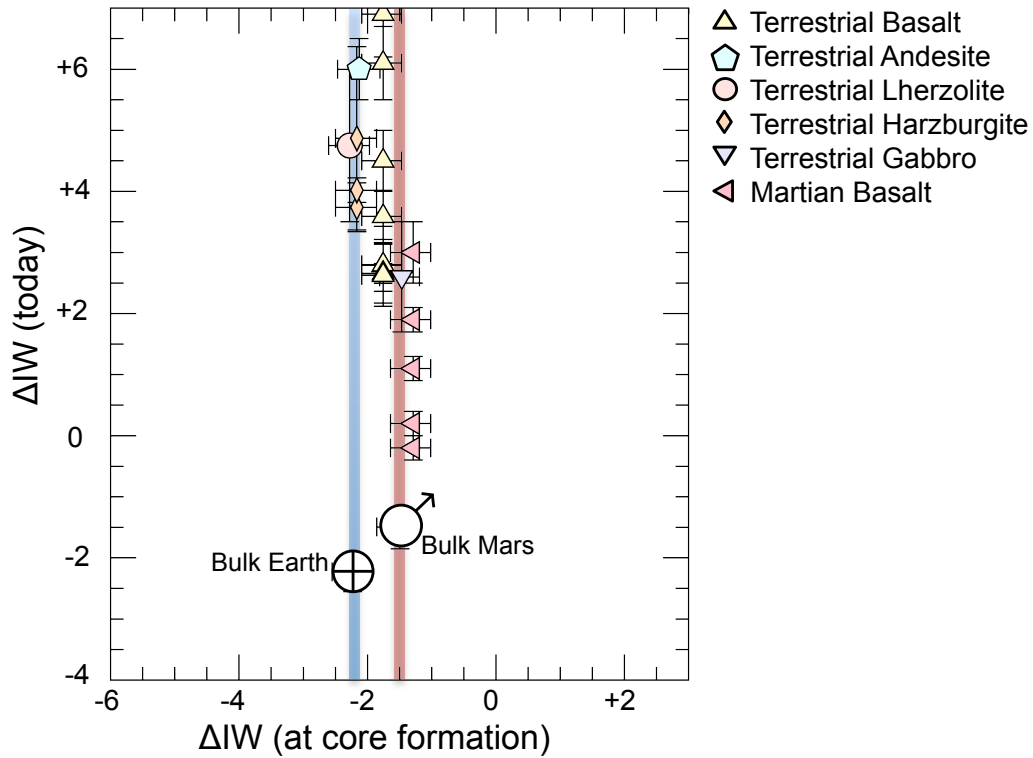

**Fig. 2. Oxygen fugacities relative to IW at core formation vs. today.** Terrestrial and martian rocks are characterized by ΔIW at the time of core formation, as calculated from the concentration of FeO, and ΔIW as measured today using various other measures of oxygen fugacity (*16*). Bulk Earth and bulk Mars values are also shown, demonstrating their similarity in intrinsic ΔIW at the time of core formation, despite the ranges in ΔIW as measured today. Error bars are 1σ (*16*). Where oxygen fugacities were previously reported relative to the quartz-fayalite-magnetite buffer (QFM), we converted them using ΔIW = ΔQFM−4. Andesite, basalt and gabbro represent crustal rocks while lherzolite and harzburgite, specific types of peridotites, are mantle rocks.

The solar protoplanetary disk must on average have had the same composition as the Sun (Supplementary Text). The oxygen fugacity of a gas with solar composition is determined by its $H_2O/H_2$ ratio, after correcting for the oxygen bound to carbon in CO and other less abundant oxides, according to the reaction $H_2 + \frac{1}{2} O_2 \rightleftharpoons H_2O$ ([16]). Studies of meteorites reveal that, like the Earth, most rocky bodies in the Solar System formed with $\Delta$IW approximately five orders of magnitude higher than that of a solar gas ([25, 26]) (Fig. 3). The presence of large amounts of iron bonded to oxygen in silicates in chondrite meteorites indicates there was a relatively high oxygen fugacity during the earliest stages of rock formation in the Solar System ([27]). The enhancement in oxygen fugacity during rocky body formation may be attributable to the sublimation of water-rich and/or rock-rich dust at high dust/gas ratios ([28]). In this context, we examine whether the processes that led to oxidation of rocks in the Solar System are typical of other planetary systems, and therefore whether the geophysical and geochemical characteristics of Earth are likely to be common among rocky exoplanets.

When the six major rock-forming elements are measured in a polluted WD, the abundance of FeO may be used to determine the oxidation state of the accreted exoplanetary rocky bodies ([2, 8, 29]). Polluted WDs with observed abundances of O, Mg, Si, Fe, Al, and Ca can be used to calculate oxygen fugacities from Equation (1) by recognizing that any Fe not bonded to oxygen must have existed as metal in the accreted bodies. Data for polluted WDs are preferable to elemental abundances in other stars due to the rocky provenance of the accreted elements in WDs, especially in the case of oxygen.

Our basic methodology is as follows: the oxide components $MgO$, $SiO_2$, $FeO$, $Al_2O_3$ and $CaO$ describe the compositions of the major minerals comprising the accreting rocks. By assigning oxygen firstly to Mg, then Si, Al, Ca and finally, Fe, we calculate the relative amount of oxidized Fe, as FeO, and assign any remaining Fe to metal representing the core of the body (*2, 8, 30*). We propagate measurement uncertainties for the polluted WDs using a Monte Carlo bootstrap approach (*16*).

We validated our method using Solar System bodies by converting the composition of these bodies into hypothetical polluted white dwarfs, as if rocks from the bodies (e.g., Earth, Mars, Mercury) had accreted onto a WD. We used typical WD measurement uncertainties for these calculations and recovered the known intrinsic oxygen fugacities for Earth, Mars, Mercury, Vesta and various chondritic bodies (*16*). The Solar System bodies span a range in ΔIW of ~6 dex, in agreement with previous studies showing that Mercury and enstatite meteorites have $f_{O_2}$ orders of magnitude lower than those for Earth, Mars, and other chondrite group meteorites (*31, 32*).

The six WDs in this study were chosen because quantitative measurements of all six major rock-forming elements are available for each. These WDs also exhibit infrared excesses, indicative of surrounding debris disks (*6*). The ΔIW values we obtain for the rocks accreted by the polluted WDs are all similar to those of Earth, Mars, Vesta, and the asteroids represented by carbonaceous (C) and ordinary (O) chondrites in the Solar System (Fig. 3).

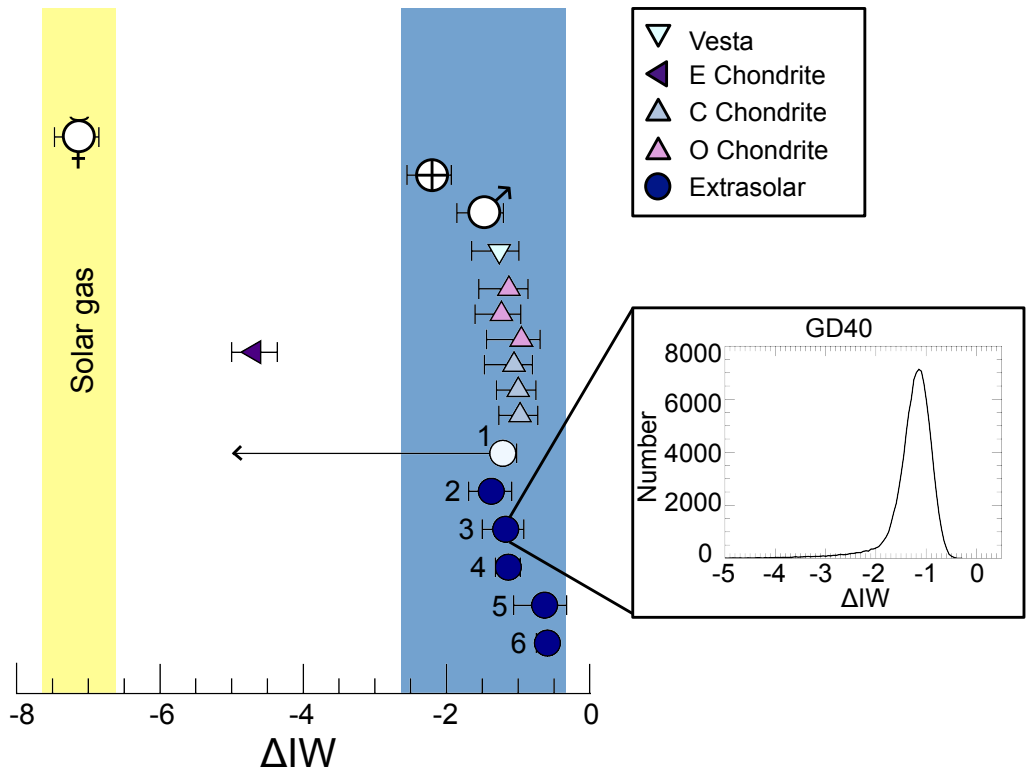

**Fig. 3. Calculated oxygen fugacities relative to IW for rocky extrasolar bodies.** Numbered circles show the values for rocky debris that polluted the white dwarfs: 1) SDSS J1043+0855; 2) WD 1536+520; 3) GD 40; 4) SDSS J0738+1835; 5) WD 1226+110; and 6) WD 1145+017. Values are listed in Table S1. One-standard-deviation error bars are from propagation of measurement uncertainties (*16*). Only an upper limit could be obtained for SDSS J1043+0855 due to the measurement uncertainties relative to the Fe concentration for that star (*16*). The ranges of relative oxygen fugacities for a gas of solar composition (yellow) and for most Solar System rocky bodies (blue) are shown for comparison. Rocks from Solar System planets are also shown and are represented by their planet symbols: Earth (⊕), Mars (♂) and Mercury (☿). Triangles show values for meteorites, representing bodies in the asteroid belt, including Vesta. The inset shows an example ΔIW probability distribution for one of the WDs, GD 40 (*16*); equivalents for the other WDs are shown in Fig. S3.

In cases in which there is both more oxygen than required to oxidize all other major elements and a commensurate amount of hydrogen, water in the accreting body is implied and the partitioning of oxygen between ice and Fe can be ambiguous (*33, 34*). In addition, unaccounted for Si in metal cores may have liberated oxygen to oxidize Fe in the accreted bodies. We find that these effects are small for the WDs in this study and do not impact the elemental abundances we used to derive oxygen fugacities (*16*) (Fig. S5).

The high oxygen fugacities of these extrasolar rocks, relative to a solar gas, suggests that whatever process oxidized rock-forming materials in the Solar System also operated in these other planetary systems. The large amount of oxidized iron in chondrite meteorites shows that oxidation relative to a solar gas occurred early in the Solar System, evidently prior to, or during the earliest stages of, planetesimal formation. Raising ΔIW by 5 log units, from solar to rock-like values, requires the gas to acquire an $H_2O/H_2$ ratio ~ 400 times that of a solar gas (*28*). This enrichment factor is greater than can be explained by simply transporting water in the form of ice particles from the outer to the inner Solar System (*28*). If dust/gas ratios control the oxidation states during rock formation, we conclude that the Solar System and the planetary systems around these six polluted WDs had similar ratios. This implies that high dust/gas ratios are intrinsic to rock formation in protoplanetary disks. A similar compositional link between planet formation in the Solar System and that around other stars is indicated by the depletion of carbon in both solar and extrasolar rocks (*35, 36*).

The high oxidation state of these rocks determined the mineralogy, and therefore the geophysical behavior, of their parent bodies or the planets these bodies formed. For example, the lower mantle of the Earth is composed of ~70% bridgmanite and ~20% magnesiowüstite, two mineral phases with markedly different rheological properties whose abundances depend on $f_{O_2}$. The relative abundances of these minerals determine the dynamic behavior of the mantle (*37*). For illustration, the influence of oxygen fugacity on the mineralogy of a silicate mantle, and the composition of a metal-rich core, can be described by the reaction

$$\text{MgO}^{\text{magnesiowüstite}} + \text{Si}^{\text{core}} + \text{O}_2 \rightleftarrows \text{MgSiO}_3{}^{\text{bridgmanite}}, \tag{3}$$

where MgO refers to the Mg component in mantle magnesiowüstite ((Mg,Fe)O), MgSiO₃ refers to the Mg component (bridgmanite) in mantle silicate perovskite ((Mg,Fe)SiO₃), and Si$^{\text{core}}$ refers to Si in the metal-rich core. Rearranging the equilibrium constant for the reaction in Equation (3), $k_{\text{eq (2)}}$, shows that the activity ratio of bridgmanite to magnesiowüstite is expected to vary with oxygen fugacity:

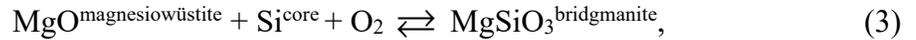

$$f_{O_2} = \frac{a_{\text{MgSiO}_3}^{\text{bridgmanite}}}{a_{\text{MgO}}^{\text{magnesiowustite}} \, a_{\text{Si}}^{\text{core}} \, k_{\text{eq(2)}}} \quad . \tag{4}$$

Equation (4) also illustrates that the Si content of the core varies inversely with oxygen fugacity. The concentrations of Si and other light elements in the core likely play a role in driving the compositional convection within the core that powers Earth's magnetic field (*38, 39*) which impacts a planet's habitability (*40*). The relative size of the metallic core of a body (or even its existence) is also determined by oxygen fugacity (*41*). If the body or bodies that accreted onto WD 1536+520 were otherwise similar to Earth or its antecedents, the ΔIW value of −1.37 would result in a planet with an Fe-rich metal core comprising ~20% by mass of the parent body. For comparison, the most highly oxidized bodies we found, with $\Delta IW \sim -0.6$, would assemble to form a planet with no Fe-rich metal core.

Our results show that the parent objects that polluted these WDs had intrinsic oxidation states similar to those of rocks in the Solar System. Based on estimates of their mass, the bodies accreting onto WDs were either asteroids that represent the building blocks of rocky exoplanets, or they were fragments of rocky exoplanets themselves (*15, 42*). In either case, our results constrain the intrinsic oxygen fugacities of rocky bodies that orbited the progenitor star of their host WD. Our data indicate that rocky exoplanets constructed from these planetesimals should be geophysically and geochemically similar to rocky planets in the Solar System, including Earth.

**Acknowledgments:** This paper benefited from the comments provided by three anonymous reviewers. **Funding:** E.D.Y. acknowledges support from the NASA Exoplanets program grant no. NNX16AB53G. A.E.D. acknowledges financial support from NASA Space Grant. H.E.S. gratefully acknowledges support from the National Aeronautics and Space Administration under grant No. 17 NAI18_2-0029 issued through the NExSS Program. **Author contributions:** A.E.D. performed the calculations and co-wrote the manuscript. E.D.Y. conceived of the project, supervised the calculations, and co-wrote the manuscript. B.Z. and B.K. analyzed the WD data. H.E.S. contributed to the statistical analysis. All authors contributed to the interpretation of the data and the preparation of the paper. **Competing interests:** The authors declare no competing interests. **Data and materials availability:** Input compositional data are available in Data S1 and our derived oxygen fugacities are listed in Table S1. The bootstrap and analytical data analysis codes are available as Data S2 and S3.

**Supplementary Materials:**
Materials and Methods
Supplementary Text
Figs S1-S6
Table S1
References (43 – 69)
Data S1
Data S2
Data S3

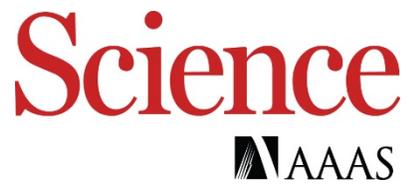

Supplementary Materials for

Oxygen fugacities of extrasolar rocks: evidence for an Earth-like geochemistry of exoplanets

Alexandra E. Doyle, Edward D. Young, Beth Klein, Ben Zuckerman, Hilke E. Schlichting.

Correspondence to: a.doyle@ucla.edu or eyoung@epss.ucla.edu

**This PDF file includes:**

## Materials and Methods

**Sources for Solar System materials in Fig. 1.** The Solar System objects used in Fig. 1 include the Sun (*43*), bulk Earth and bulk silicate Earth (*44*), bulk Mars (*45*), bulk Mercury (*46, 47*), the Moon (*48*), Vesta (*49*), Comet Halley (*50*), terrestrial mid-ocean ridge basalts (MORBs) (*51*), chondrites (CI (*43*), L, and EH-EL silicate (*52*)), and terrestrial lherzolite (*53*).

**Calculation of oxygen fugacities and associated uncertainties from white dwarf element ratio data.** Here we provide a step-by-step summary of our calculation method for obtaining oxygen fugacities from the white dwarf elemental ratios. We include analytical estimates for the uncertainties at each step using the general error propagation equation (*54*) based on linear expansion of the contributing variances. The equation is a basic method for transforming variances in independent variables $u$, $v$, etc., represented by $\sigma_u$ and $\sigma_v$, respectively, to that for dependent variable $x$ where $x = f(u,v,\ldots)$:

$$\sigma_x^2 = \left(\frac{\partial x}{\partial u}\right)^2 \sigma_u^2 + \left(\frac{\partial x}{\partial v}\right)^2 \sigma_v^2 + \cdots . \tag{S1}$$

Although useful as a guide to the sources of uncertainty in the calculations, these analytical estimates of uncertainty are complemented by the bootstrap error analysis described below.

The input data for calculating oxygen fugacities are values for $\log(Z/X)$ where $Z$ represents the atomic abundances of metals O, Si, Mg, Ca, Al, and Fe and $X$ is the atomic abundance of either H or He, depending on the WD type in question (WD types are defined below). Uncertainties in these logs of ratios are usually reported as symmetrical errors in logarithmic space (e.g., $-4.0 \pm 0.3$ dex), and a lognormal distribution for $Z/X$ is adopted here.

**Step 1**: We first use the $\log(Z/X)$ values to derive $\log(Z/O)$ where

$$\log(Z / O) = \log(Z / X) - \log(O / X) . \tag{S2}$$

The associated uncertainties prescribed by Equation (S1) are:

$$\sigma_{\log(Z/O)} = \sqrt{\sigma_{\log(Z/X)}^2 + \sigma_{\log(O/X)}^2} . \tag{S3}$$

**Step 2**: Next we convert $\log(Z/O)$ to $Z/O$ (using $Z/O = 10^{\log(Z/O)}$) and use Equation (S1) to derive the uncertainty in the ratios:

$$\sigma_{Z/O}^2 = \left(\frac{\partial(Z / O)}{\partial \log(Z / O)}\right)^2 \sigma_{\log(Z/O)}^2 . \tag{S4}$$

Because

$$\frac{\partial (Z/O)}{\partial \log(Z/O)} = \frac{\partial (Z/O)}{\partial \ln(Z/O)} \frac{\partial \ln(Z/O)}{\partial \log(Z/O)}$$
$$= (Z/O) \frac{\partial \ln(Z/O)}{\partial \log(Z/O)}$$

(S5)

and the change of base is afforded by $\ln(x) = \ln(10) \log(x)$ so that $d\ln(x)/d\log(x) = \ln(10)$, we arrive at

$$\sigma_{Z/O} = (Z/O) \ln(10) \sigma_{\log(Z/O)} \ .$$

(S6)

This expression identifies the errors in $\log(Z/O)$ as relative errors in the $Z/O$ ratios because

$$\frac{\sigma_{Z/O}}{(Z/O)} = \ln(10) \sigma_{\log(Z/O)} \ .$$

(S7)

Equation (S7) yields the uncertainties in $Z/O$ that can be propagated through the remainder of the calculation.

**Step 3**:  The oxygen atoms are assigned to Si, Mg, Al, and Ca to form the oxides $SiO_2$, $MgO$, $Al_2O_3$, and $CaO$.  The relationship between the ratios of $Z/O$ and the number of oxygens required by each oxide $i$, $(O)_i$, are

$$\frac{(O)_{SiO_2}}{O_{Total}} = 2 \frac{Si}{O_{Total}}$$

$$\frac{(O)_{MgO}}{O_{Total}} = \frac{Mg}{O_{Total}}$$

$$\frac{(O)_{Al_2O_3}}{O_{Total}} = (3/2) \frac{Al}{O_{Total}}$$

$$\frac{(O)_{CaO}}{O_{Total}} = \frac{Ca}{O_{Total}}$$

(S8)

where from here forward we make the distinction between oxygen assigned to oxide $i$ and total oxygen, $O_{Total}$, and $(O)_i/O_{Total}$ is the ratio of the number of oxygen atoms comprising oxide $i$ to the total number of oxygen atoms.  The excess oxygen $(O_{XS})$ available for bonding to Fe as FeO relative to the total oxygen is calculated as

$$\frac{O_{XS}}{O_{Total}} = 1 - \frac{(O)_{SiO_2}}{O_{Total}} - \frac{(O)_{MgO}}{O_{Total}} - \frac{(O)_{Al_2O_3}}{O_{Total}} - \frac{(O)_{CaO}}{O_{Total}} \ , \tag{S9}$$

where the 1 represents $O_{Total}/O_{Total}$. Applying Equation (S1) to calculate the uncertainty in $O_{XS}/O_{Total}$ using the uncertainties in the $Z/O$ ratios from Equations (S4-S7) gives

$$\sigma_{O_{XS}/O_{Total}} = \sqrt{4\sigma^2_{Si/O_{Total}} + \sigma^2_{Mg/O_{Total}} + (3/2)^2 \sigma^2_{Al/O_{Total}} + \sigma^2_{Ca/O_{Total}}} \tag{S10}$$

**Step 4**: The excess oxygen, $O_{XS}$, is assigned to Fe as FeO. If $Fe/O_{Total} > O_{XS}/O_{Total}$, $FeO/O_{Total} = O_{XS}/O_{Total}$. Conversely, if $Fe/O_{Total} < O_{XS}/O_{Total}$, $FeO/O_{Total} = Fe/O_{Total}$. More formally,

$$\frac{FeO}{O_{Total}} = \begin{cases} \dfrac{O_{XS}}{O_{Total}} & \text{if } Fe/O_{Total} > O_{XS}/O_{Total} \\[2mm] \dfrac{Fe}{O_{Total}} & \text{if } Fe/O_{Total} \leq O_{XS}/O_{Total} \end{cases} . \tag{S11}$$

Where oxygen is overabundant relative to the other elements, the uncertainty in $FeO/O_{Total} = \sigma_{Fe/O}$ given by Equation (S6). Where O is less than Fe, uncertainty in $FeO/O_{Total}$ is the uncertainty in excess oxygen given by Equation (S10).

**Step 5**: The oxide ratios relative to total oxygen are used to calculate the mole fraction of FeO for the accreted material in the white dwarf. The ratios of oxides to total oxygen in addition to $FeO/O_{Total}$, given in Equation (S11), are

$$\begin{aligned} \frac{SiO_2}{O_{Total}} &= \frac{Si}{O_{Total}} \\[3mm] \frac{MgO}{O_{Total}} &= \frac{Mg}{O_{Total}} \\[3mm] \frac{Al_2O_3}{O_{Total}} &= 0.5\frac{Al}{O_{Total}} \\[3mm] \frac{CaO}{O_{Total}} &= \frac{Ca}{O_{Total}} \end{aligned} \tag{S12}$$

The mole fraction of FeO, $x_{FeO}$, is calculated as

$$x_{\text{FeO}} = \frac{\text{FeO/O}_{\text{Total}}}{\text{FeO/O}_{\text{Total}} + \text{SiO}_2/\text{O}_{\text{Total}} + \text{MgO/O}_{\text{Total}} + \text{Al}_2\text{O}_3/\text{O}_{\text{Total}} + \text{CaO/O}_{\text{Total}}} \ . \qquad \text{(S13)}$$

The uncertainty in the mole fraction of FeO can be estimated using Equation (S1) applied to Equation (S13):

$$\sigma^2_{x_{\text{FeO}}} = \left(\frac{\partial x_{\text{FeO}}}{\partial \text{FeO/O}_{\text{Total}}}\right)^2 \sigma^2_{\text{FeO/O}_{\text{Total}}} + \left(\frac{\partial x_{\text{FeO}}}{\partial \text{SiO}_2/\text{O}_{\text{Total}}}\right)^2 \sigma^2_{\text{SiO}_2/\text{O}_{\text{Total}}} + \left(\frac{\partial x_{\text{FeO}}}{\partial \text{MgO/O}_{\text{Total}}}\right)^2 \sigma^2_{\text{MgO}_{\text{Total}}} +$$
$$\left(\frac{\partial x_{\text{FeO}}}{\partial \text{Al}_2\text{O}_3/\text{O}_{\text{Total}}}\right)^2 \sigma^2_{\text{Al}_2\text{O}_3/\text{O}_{\text{Total}}} + \left(\frac{\partial x_{\text{FeO}}}{\partial \text{CaO/O}_{\text{Total}}}\right)^2 \sigma^2_{\text{CaO/O}_{\text{Total}}} \qquad \text{.(S14)}$$

Evaluating the partial derivatives in Equation (S14) provides the analytical estimate for the uncertainty in $x_{\text{FeO}}$:

$$\sigma^2_{x_{\text{FeO}}} = \left(\frac{\frac{\text{SiO}_2}{\text{O}_{\text{Total}}} + \frac{\text{MgO}}{\text{O}_{\text{Total}}} + \frac{\text{Al}_2\text{O}_3}{\text{O}_{\text{Total}}} + \frac{\text{CaO}}{\text{O}_{\text{Total}}}}{\left(\frac{\text{FeO}}{\text{O}_{\text{Total}}} + \frac{\text{SiO}_2}{\text{O}_{\text{Total}}} + \frac{\text{MgO}}{\text{O}_{\text{Total}}} + \frac{\text{Al}_2\text{O}_3}{\text{O}_{\text{Total}}} + \frac{\text{CaO}}{\text{O}_{\text{Total}}}\right)^2}\right)^2 \sigma^2_{\text{FeO/O}_{\text{Total}}} +$$

$$\sum_{i \neq \text{FeO}} \left(\frac{-\frac{(Z_x O_y)_i}{\text{O}_{\text{Total}}}}{\left(\frac{\text{FeO}}{\text{O}_{\text{Total}}} + \frac{\text{SiO}_2}{\text{O}_{\text{Total}}} + \frac{\text{MgO}}{\text{O}_{\text{Total}}} + \frac{\text{Al}_2\text{O}_3}{\text{O}_{\text{Total}}} + \frac{\text{CaO}}{\text{O}_{\text{Total}}}\right)^2}\right)^2 \sigma^2_{(Z_x O_y)i/\text{O}_{\text{Total}}} \ , \qquad \text{(S15)}$$

where $(Z_x O_y)_i$ represents the oxides other than FeO.

**Step 6**: The final step for obtaining a single value for the oxygen fugacity expressed as $\Delta$IW is to use the values of $x_{\text{FeO}}$ according to

$$\Delta\text{IW} = 2\log\left(\frac{x_{\text{FeO}}}{0.85}\right) \qquad \text{(S16)}$$

where the value 0.85 represents a nominal mole fraction of Fe in the metal phase during the formation of the accreting body. Again, using Equation (S1) applied to Equation (S15) an estimate for the uncertainty in $\Delta$IW can be obtained:

$$\sigma^2_{\Delta\text{IW}} = \left(\frac{2}{\ln(10)x_{\text{FeO}}}\right)^2 \sigma^2_{x_{\text{FeO}}} \qquad \text{(S17)}$$

Application of these six steps together with the estimates of uncertainties applied to the data for the six polluted white dwarfs in this study are provided in Data S1. In practice, Equation (S17) does not capture the asymmetry in $\Delta$IW uncertainties, resulting in unrealistic lower bounds when $\Delta$IW values are low.

When Fe > $O_{XS}$, FeO/$O_{Total}$ in the accreted rock is equated with $O_{XS}$/$O_{Total}$ and the uncertainties in this parameter are large due to the combined errors from all of the elements analyzed (Equation S10). The large uncertainties can preclude the determination of accurate $\Delta$IW values. This situation arises for one white dwarf in this study. As a result, only an upper limit on $\Delta$IW could be obtained for SDSS J1043+0855, which we demonstrated with the bootstrap method described below.

**Bootstrap Analysis**. Although estimates of uncertainties can be obtained in the majority of polluted white dwarfs and simulated white dwarfs in this study by simply using Equation (S1), the analytical estimates assume symmetrical uncertainties about the mean, so are insufficient where the mole fractions of FeO ($x_{\text{FeO}}$) are so low that these uncertainties can lead to values of zero. We therefore use a bootstrap method with replacement (55) to estimate median $\Delta$IW values and the uncertainties about the medians. The bootstrap approach also captures the effects of non-Gaussian probability distributions that arise when transforming the data using steps 1 through 6 described above.

The bootstrap procedure carries out steps 1 through 6 where the analytical error propagation expressions (Equations S3, S4-S7, S10, S14, S15, and S17) are replaced by Monte Carlo sampling. The method is summarized in Fig. S1. Our input data are log($Z/X$) values where $X$ is He or H and $Z$ is the element of interest. We use normal distributions for log($Z/X$) as dictated by the symmetrical uncertainties reported for the white dwarf abundance ratios. The resulting lognormal distributions ensure positive element ratios throughout the error propagation.

We draw at random single values for log($Z/X$) (e.g., log(Mg/He), log(Si/He), …) from the normal distributions defined by their means and uncertainties as a representation of the single measurement available for each white dwarf. We then calculate $\Delta$IW from this single draw, and repeat this calculation 100,000 times to obtain a frequency distribution of $\Delta$IW values.

The resulting frequency distributions of 100,000 $\Delta$IW values are asymmetric (Fig. S1). We therefore report median $\Delta$IW values and calculate asymmetrical uncertainties based on an extension of the interquartile range, centered on the median, to encompass 67% of the distribution rather than just 50%. This procedure makes our reported uncertainties equivalent to the 1$\sigma$ error bars commonly reported for normal distributions.

We compare the results from our study using the bootstrap method with the results from the analytical calculations in Fig. S2. The results summarized in Fig. S2 show that the $\Delta$IW values for the WDs and Solar System rocks (in fictive white dwarfs) obtained from the analytical and the bootstrap methods are identical (one white dwarf value is actually just an upper limit). However, the uncertainties differ for objects with low $x_{\text{FeO}}$.

For the white dwarf SDSS J1043+0855 we are only able to obtain an upper limit to the $\Delta$IW value. The data for this white dwarf exhibit approximately twice the uncertainty in the rock-forming elements (Fe, Si) compared with those for the other five white dwarfs. While our bootstrap method based on lognormal $Z/X$ yields only positive element ratios, the accumulation of large relative uncertainties associated with these $Z/X$ values in this case leads to insufficient oxygen to accommodate MgO + SiO$_2$ + Al$_2$O$_3$ + CaO (Equations S9 and S10) in a large fraction of the random draws. In these cases, there is no oxygen available to pair with iron to make FeO and ΔIW values cannot be calculated. The implication for these draws in which FeO is not present is that the associated ΔIW values are exceedingly low (lower than that for Mercury, for example). However, because instances of $FeO/O_{Total} = 0$ cannot be included in the calculation of ΔIW values (ΔIW approaches negative infinity in these instances), the ΔIW values obtained from only the random draws where FeO is present bias the results. Therefore, we can only assign an upper limit based on the 83.5 percentile for ΔIW for this single white dwarf. Frequency distributions for ΔIW values for the remaining five of the WDs in this study are shown in Fig. S3.

**The definition of ΔIW**. The ΔIW parameter is a standard method for removing the influences of differences in both temperature ($T$) and pressure ($P$) on reported log($f_{O_2}$) values, allowing direct comparisons of oxygen fugacities among materials. We present here the thermodynamic basis for the ΔIW units used in the paper, showing why $T$ and $P$ are effectively removed as variables when reporting $f_{O_2}$ using this reference frame.

The reaction between metallic iron (Fe) and wüstite (FeO) that provides our reference oxygen fugacity is

$$Fe + \frac{1}{2}O_2 \rightleftharpoons FeO \ . \tag{S18}$$

Thermodynamic chemical equilibrium for this reference reaction is described by the sum of chemical potentials for each species multiplied by the reaction stoichiometric coefficients, resulting in

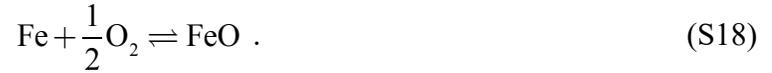

$$0 = \Delta\hat{H}^{\circ}_{rxn} - T\Delta\hat{S}^{\circ}_{rxn} + \Delta\hat{V}^{\circ}_{solids}(P - P^{\circ}) + RT\ln(a_{FeO}) - RT\ln(a_{Fe}^{metal}) - \frac{1}{2}RT\ln(f_{O_2}) \tag{S19}$$

where the molar enthalpy of reaction ($\Delta\hat{H}^{\circ}_{rxn}$), the molar entropy of reaction ($\Delta\hat{S}^{\circ}_{rxn}$), and the molar volume change for solids in the reaction ($\Delta\hat{V}^{\circ}_{solids}$) refer to the values at standard state conditions (as indicated by the ° superscript), taken to be standard pressure ($P^{\circ}$, usually 1 bar), compositionally pure phases, and the temperature of interest. The terms $a_{FeO}$ and $a_{Fe}^{metal}$ are the activities of the indicated components in their respective phases that correct for impure phase compositions, and $R$ is the ideal gas constant. The oxygen fugacity includes the activity for oxygen in an impure gas as well as the pressure of gas relative to standard state. It is convenient to collect the activity terms, including the oxygen fugacity, by rearranging to arrive at

$$\ln(f_{O_2}) - 2\ln(a_{FeO}) + 2\ln(a_{Fe}^{metal}) = 2\frac{\left[\Delta\hat{H}^{\circ}_{rxn} - T\Delta\hat{S}^{\circ}_{rxn} + \Delta\hat{V}^{\circ}_{solids}(P - P^{\circ})\right]}{RT} \ . \tag{S20}$$

This expression shows explicitly the relationship between oxygen fugacity, the activities of the components in the solids, temperature, and pressure. In the case of pure iron metal and pure FeO, our IW reference, the thermodynamic expression for equilibrium reduces to

$$\ln(f_{O_2})_{IW} = 2\frac{\left[\Delta\hat{H}^\circ_{rxn} - T\Delta\hat{S}^\circ_{rxn} + \Delta\hat{V}^\circ_{solids}(P - P^\circ)\right]}{RT} \tag{S21}$$

because $a_{FeO} = a_{Fe}^{metal} = 1$ where the solid phases are pure. We define $\Delta IW$ as the difference between Equation (S20) for the general case and Equation (S21) for pure iron and wüstite, IW, yielding:

$$\ln(f_{O_2}) - 2\ln(a_{FeO}) + 2\ln(a_{Fe}^{metal}) - \ln(f_{O_2})_{IW}$$

$$= 2\frac{\left[\Delta\hat{H}^\circ_{rxn} - T\Delta\hat{S}^\circ_{rxn} + \Delta\hat{V}^\circ_{solids}(P - P^\circ)\right]}{RT} - 2\frac{\left[\Delta\hat{H}^\circ_{rxn} - T\Delta\hat{S}^\circ_{rxn} + \Delta\hat{V}^\circ_{solids}(P - P^\circ)\right]}{RT} \tag{S22}$$

$$= 0$$

which, using the definition of $\Delta IW$, reduces to

$$\Delta IW = \log(f_{O_2}) - \log(f_{O_2})_{IW} = 2\log\left(\frac{a_{FeO}}{a_{Fe}^{metal}}\right),$$

$$= 2\log\left(\frac{x_{FeO}}{x_{Fe}^{metal}}\right) + 2\log\left(\frac{\gamma_{FeO}}{\gamma_{Fe}^{metal}}\right) \tag{S23}$$

where we have converted from natural log to base 10, $x_i$ are mole fractions, and $\gamma_i$ are activity coefficients (and making use of $a_i = x_i \gamma_i$). By taking the difference between Equations (S20) and (S21), the temperature and pressure-dependent terms of the thermodynamic expressions, the right-hand sides of both equations, cancel (Equation S22). This shows that while oxygen fugacity depends on both $T$ and $P$, this dependency is removed by using a reference reaction, in this case the IW equilibrium. By applying this formalism, we relate the activity of FeO in a silicate melt, for example, to a standard state of pure FeO (the wüstite mineral phase) (*56*). Differences in the behavior of the FeO component from host phase to host phase will therefore be embodied in the activity coefficients in Equation (S6). For simplicity, and because the values range from about 1 to as high as ~4, we ignore activity coefficients in our calculations (a common assumption where activity coefficients are not well constrained), adding an ambiguity in the exact oxygen fugacities of up to ~ ½ dex. In practice, the potential inaccuracy is less than this because activity coefficients are not likely to be disparate by factors of four between the different FeO-bearing minerals of interest.

This summary of the thermodynamic basis for reporting oxygen fugacities as $\Delta IW$ values shows why no temperature or pressure corrections are necessary to first order. This is not to say that oxygen fugacity does not change with pressure within a planet (*57, 58*), but these variations, e.g., Fig. 2, are secondary to what we refer to as the intrinsic oxygen fugacities defined by the overall FeO and metal contents of the bodies.

**Validating the Method.** The rocky, Solar System objects used in this study include bulk silicate Earth (*44*), bulk silicate Mars (*45*), bulk silicate Mercury (*46*), Vesta (*49*), terrestrial mid-ocean ridge basalts (MORBs) (*51*), chondrites (CI (*43*), CM, CV, H, L, LL and EH-EL silicate (*52*)), terrestrial gabbro (*59*), terrestrial harzburgite (*59*), terrestrial lherzolite (*53*), terrestrial andesite (*60*) and martian basalt (*61*). We converted weight percentages to abundances by number and scaled $Z$/He for each element $Z$ to the Si/He ratio for GD 40 in order to simulate WD stars polluted with these objects. We then gave the log($Z$/He) values errors equal to the uncertainties for GD 40 (*8, 12*). The results for these Solar System bodies span a range in ΔIW of ~6 dex, and demonstrate that we can detect accurate ΔIW values from ~ −8 to ~ −0.9 range with our method (Fig. S4).

**The Effects of Water and Si in Metal Cores.** Where there is substantial hydrogen in the atmosphere of the polluted white dwarfs, ambiguity arises regarding how much oxygen was bound as ices versus how much oxygen bonded with Fe to form silicates and oxides (the "FeO" component); unmelted ice and metal could have coexisted. An inability to disentangle this difference excludes these white dwarfs from our analysis. However, previous literature for the six WDs in this study, using similar methods for budgeting oxygen, have concluded that, at most, there could have been 10% to 20% water by mass. For reference, Ceres is inferred to be ~30% water by mass (*62*). If we assign 10% of the calculated minimum mass accreted onto GD 40 to $H_2O$, and remove that oxygen so that it cannot be used to oxidize other elements, the ΔIW value for GD 40 shifts by only ~0.01 (Fig. S5). Thus, the inclusion of the possible presence of water does not strongly affect our results.

For the purposes of our calculation, we assume 0 wt. % Si in the cores of these bodies, but, previous studies estimate that 2-8 wt. % Si may be present in Earth's core (*63*) and Fe-rich metals in aubrite meteorites, igneous equivalents of enstatite chondrites, contain ~2 wt. % Si (*64*). For reference, 8 wt. % Si in the core is the equivalent of having ~18% of Earth's total Si atoms in the core. In order to explore this effect, we assigned 18% of the Si atomic abundance to the core for GD 40, liberating some oxygen to oxidize Fe that would have otherwise oxidized Si in our model. In this example, ΔIW for GD 40 changes by < 0.05 (Fig. S5). Thus, the additional uncertainty in ΔIW due to plausible concentrations of Si in the metal core is small.

**Determining Element Ratios in WDs.** For the WDs used in this study, we used relative atomic abundances for the six major rock-forming elements O, Mg, Si, Fe, Al, and Ca and their respective measurement uncertainties. In cases where a range of uncertainties was given, we use the average. White dwarf cores are composed primarily of carbon and oxygen surrounded by a thin outer envelope of hydrogen and/or helium. DA and DB white dwarfs have hydrogen and helium-dominated atmospheres, respectively. The letter D stands for "degenerate" (*65, 66*). White dwarfs that display atoms heavier than helium in their atmospheres, such as those used in this study, can be classified with the letter 'Z.' The classification of WDs is therefore based on the predominance of H, He, or metals in their atmospheres (classifications are listed in Table S1).

We use ratios of elements rather than absolute abundances in order to mitigate the uncertainties surrounding the nature of the accretion process and modeling of the WD properties that affect the concentrations of the elements similarly (*67*). Polluting elements in DB WDs with temperatures similar to those in our sample set settle in $10^3$ to $10^6$ years whereas metals in DAs settle in days (*68*). Concentration ratios for the rocky accreted bodies are derived from those observed in the WDs using the equation for the time-dependent addition of element $Z$ resulting from addition to the convective layer or photosphere:

$$\frac{dM_Z}{dt} = \dot{M}_Z - \frac{M_Z}{\tau_Z} \qquad (S24)$$

where $M_Z$ is the mass of element $Z$ in the white dwarf mixing layer, $\dot{M}_Z$ is the accretion rate of the element onto the star, $t$ is the elapsed time for accretion, and $\tau_Z$ is the characteristic time for settling out of the convective layer or photosphere for element $Z$. This first-order linear differential equation has the usual general solution

$$M_Z = ce^{-t/\tau_Z} + e^{-t/\tau_Z}\int e^{t/\tau_Z}\dot{M}_Z(t)dt \qquad (S25)$$

where $c$ is an integration constant that we set to zero because the mass of $Z$ at time zero is taken to be zero. For the He-rich DB white dwarfs with temperatures < 17000K, the settling times may be long compared with accretion times (*69*) and the large values for $\tau_Z$ relative to $t$ lead to the solution

$$M_Z = \int \dot{M}_Z(t)dt \qquad (S26)$$

We consider that due to the long settling times (e.g., $\tau_Z \sim 5{\times}10^5$ years at 14000K) there is a buildup of the polluting elements in the convective layer owing to a constant rate of accretion that requires several times $\tau_Z$, corresponding to several million years, for steady-state to be attained. In this case the ratio of any two elements observed in the WD atmosphere, $Z_1$ and $Z_2$, faithfully reflects the ratio for the accreting body because for the WD:

$$\frac{M_{Z_2}}{M_{Z_1}} = \frac{\dot{M}_{Z_2}}{\dot{M}_{Z_1}} \quad . \qquad (S27)$$

Here the ratio of constant rates on the right-hand side of Equation (S27) corresponds to the element ratio in the parent body of the accreted material and the mass ratio on the left is the ratio in the mixing layer of the star.

For the two DA white dwarfs in this study, the timescale for settling will be much shorter and the elemental abundances should only be observable if they are in steady state resulting from a balance between accretion and settling rates. In these cases, exp(-$t/\tau_Z$) approaches zero, so the solution for $M_Z$ with constant accretion rate becomes

$$\begin{aligned} M_Z &= e^{-t/\tau_Z}\int e^{t/\tau_Z}\dot{M}_Z(t)dt \\ &= \dot{M}_Z\tau_Z\left(1 - e^{-t/\tau_Z}\right) \\ &= \dot{M}_Z\tau_Z \end{aligned} \qquad (S28)$$

and the observed ratio of two elements in the WD is in this case

$$\frac{M_{Z_2}}{M_{Z_1}} = \frac{\dot{M}_{Z_2}\tau_{Z_2}}{\dot{M}_{Z_1}\tau_{Z_1}} \, . \tag{S29}$$

The element ratio in the parent body of the accreting body is the ratio of rates on the right-hand-side of Equation (S29). In these circumstances the element ratio in the accreting parent body is obtained by multiplying the observed element ratio in the white dwarf, the left-hand side of (S29), by the inverse ratio of settling times of the different elements. Settling times depend on temperature and gravity of the white dwarf that are modeled. Fortunately, settling times for most rock-forming elements under a wide range of conditions are similar to one another within a factor of two (*68*). For the two DA WDs in this study, SDSS J1043+0855 and WD 1226+110, we used steady-state abundances corrected for settling times (*9, 10*). Properties of the WDs used in this study are listed in Table S1. The high temperature of DB WD 1536+520 suggests that the steady-state solution may be more appropriate than the buildup model for this star, although the differences are within uncertainties. For illustration, we show $\Delta$IW values for the DB white dwarf GD 40 based on both the buildup and steady-state solutions for elemental ratios (Fig. S5). We find that because of similarities in settling times of the rock-forming cations, observed $\Delta$IW values do not change greatly for millions of years in the WDs as long as accretion is active.

## Supplementary Text

**Comparison of $\Delta$IW between solar gas and rocky bodies.** When discussing the oxygen fugacity of a solar gas, we adopt the zero-order assumption that the gas that comprised the protoplanetary disks in which rocks initially formed are thought of as compositionally similar to the gas comprising the Sun. Such a gas will be highly reduced due to the presence of abundant $H_2$. The oxygen fugacity for a gas of solar composition is defined by the reaction $H_2 + \frac{1}{2}O_2 = H_2O$ because water is an oxygen-bearing molecule in the gas (the other major O-bearing species being CO based on kinetic and thermodynamic calculations and observations of protoplanetary disks). The oxygen fugacity defined by the equilibrium between water and $H_2$ gas can be compared directly with the oxygen fugacities recorded by the rocks to obtain a relative measure of oxidizing power. The distribution of Fe between metal and silicate or oxide during the earliest stages of mineral formation in the protoplanetary disk would have been controlled by the oxygen fugacity of the solar gas in the absence of processes that would concentrate oxygen. The latter might include evaporation or sublimation of water ice, silicate dust, or oxide dust. As an illustration of comparing oxygen fugacity in rocky materials with a solar gas, we show in Fig. S6 some representative oxygen fugacity vs. temperature curves (referred to by geochemists as oxygen fugacity "buffer" curves) typically used in geochemistry, including iron-wüstite (IW), representing the equilibrium reaction $Fe + \frac{1}{2}O_2 = FeO$, quartz-fayalite-magnetite (QFM), representing the reaction $3Fe_2SiO_4 + O_2 = 2Fe_3O_4 + 3SiO_2$, and lastly hematite-magnetite (HM), representing the reaction $2Fe_3O_4 + \frac{1}{2}O_2 = 3Fe_2O_3$. These curves were calculated from thermodynamic expressions (*70*) for an arbitrary pressure of 1 bar ($10^5$ Pa) (the curves are the same for $P \ll 1$ bar as well). For comparison we show the $\log f_{O_2}$ vs. $T$ curve for a solar gas based on thermodynamic data (*71*). Also shown is the definition of $\Delta$IW for such a gas, illustrating that it is $\sim 5$ orders of magnitude lower in $\log f_{O_2}$ than the IW reference regardless of temperature (and pressures relevant to protoplanetary disk formation). One might assume that planets grew in a reducing environment like that imposed by the hydrogen-rich disk gas with a solar-like composition, i.e., with oxygen fugacities like those defining the solar gas curve in Fig. S6. However, the fact that rocks record oxygen fugacities orders of magnitude higher than this means that an unknown process oxidized the environments in which the rocks formed relative to the hydrogen-rich solar-like gas. The higher oxygen fugacity transferred Fe from metal to oxidized forms. The precise mechanism for this enrichment is debated.

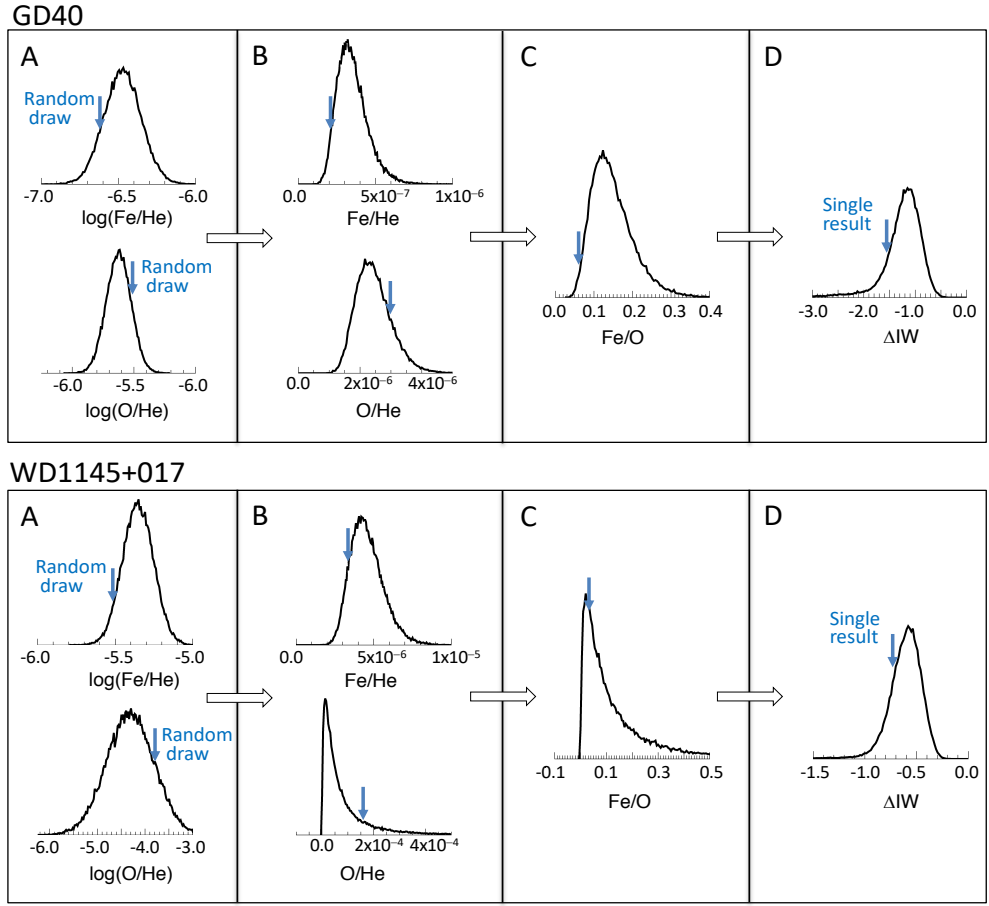

**Fig. S1. Summary diagram showing the bootstrap method parent probability distributions for two WDs.** GD 40 represents a well-behaved example, and WD 1145+017 is the WD with the largest uncertainties in O/He, leading to pronounced asymmetry in the $Z$/O distribution. 100,000 random samplings from normal distributions of log($Z$/He) comprise each probability distribution. A single random draw is shown by the blue vertical arrows. Progressing from panel A to D for each WD, the input log(Fe/He) and log(O/He) values (blue arrows marked "random draw" in A), yield the indicated values for Fe/He and O/He (B), Fe/O (C) and finally the indicated ΔIW values (D). The ΔIW distributions shown (D) include the propagation of the probability distributions for all $Z$/He values, not just Fe/He and O/He.

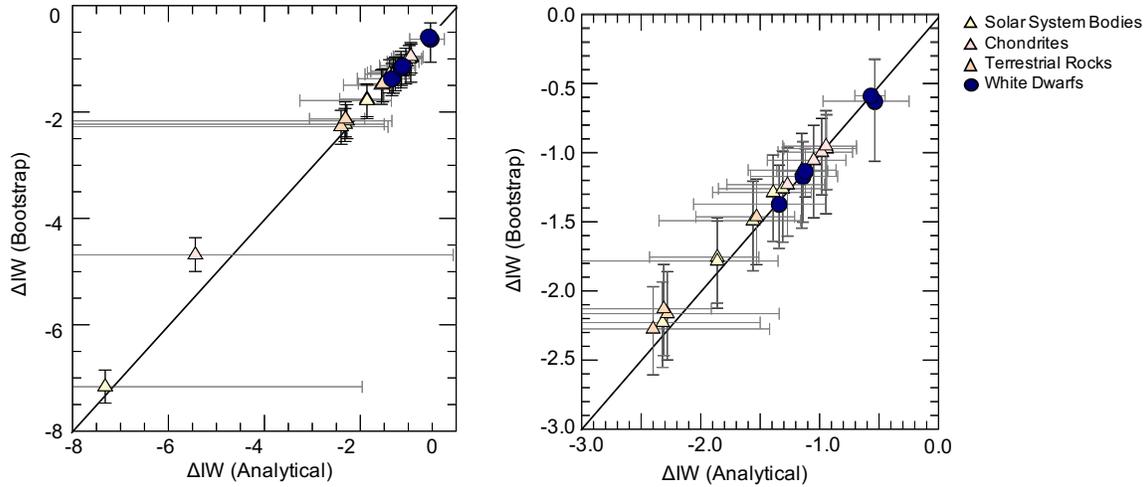

**Fig. S2. Comparisons of our two methods of calculating ΔIW values and associated uncertainties.** The simpler analytical method, Equations (S1-S17), is compared with the bootstrap method where the $Z/X$ values are assumed to be lognormally distributed. While values for ΔIW agree between the two methods, the uncertainties obtained by simple linear propagation of errors, (horizontal error bars in the panel at left) become too large at low ΔIW for typical polluted white dwarf measurement uncertainties. White dwarf data are shown as circles while triangles are Solar System rocks added to fictive white dwarfs with measurement uncertainties resembling those for GD 40. The lowest ΔIW values are Mercury and enstatite chondrite silicate in the left-hand panel. 1:1 correlation lines are shown for reference. The large low-side error bar for white dwarf SDSS J1043+0855 (white circle) reflects the fact that the calculated ΔIW value is an upper limit due to FeO being within error of zero. The right-hand panel shows a close-up of the white dwarf data compared with the Solar System test data.

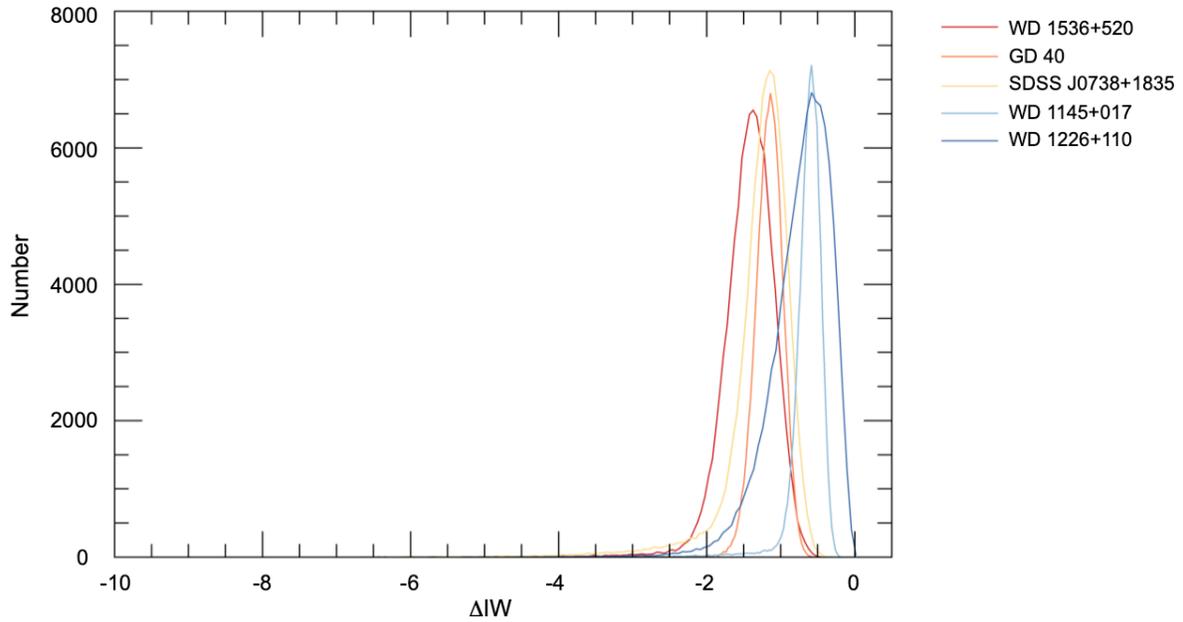

**Fig. S3. Frequency distributions from bootstrap analysis for WDs.** Frequency distributions of 100,000 ΔIW values from bootstrap samplings of log($Z/X$) (with replacement) obtained for five polluted white dwarfs. The frequency distribution for SDSS 1043+0855 cannot be shown here because the data define only an upper limit for ΔIW. The result for this WD is shown in Fig. 3.

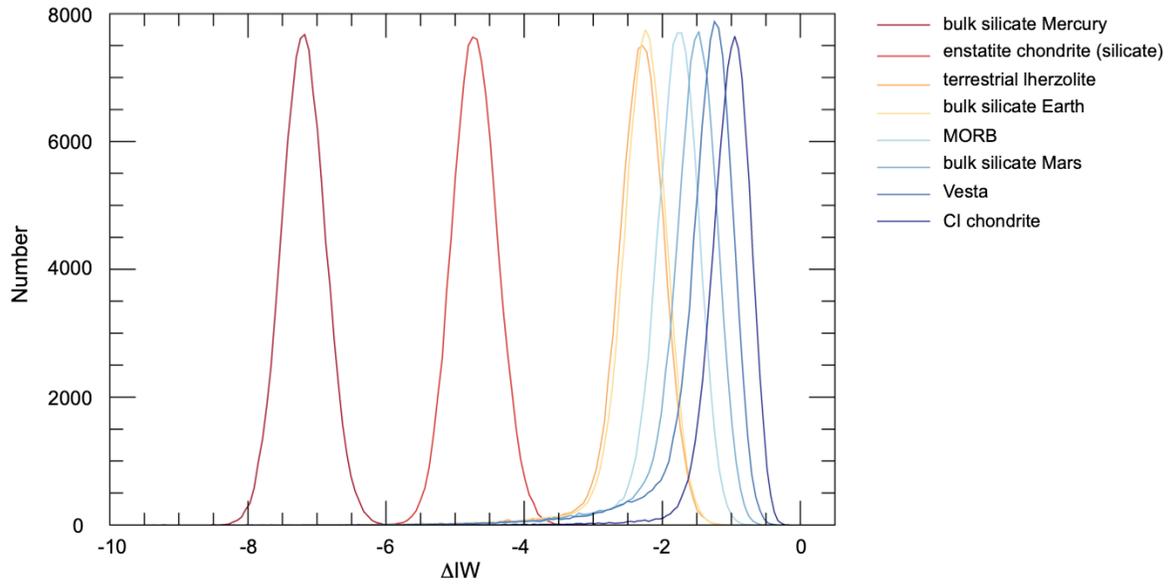

**Fig. S4. Frequency distributions from bootstrap analysis for Solar System rocks.**
Frequency distributions of 100,000 ΔIW values from bootstrap samplings of log($Z/X$)
(with replacement) obtained for Solar System rocks polluting fictive white dwarfs in
this study. Legend order corresponds to ΔIW from low to high values.

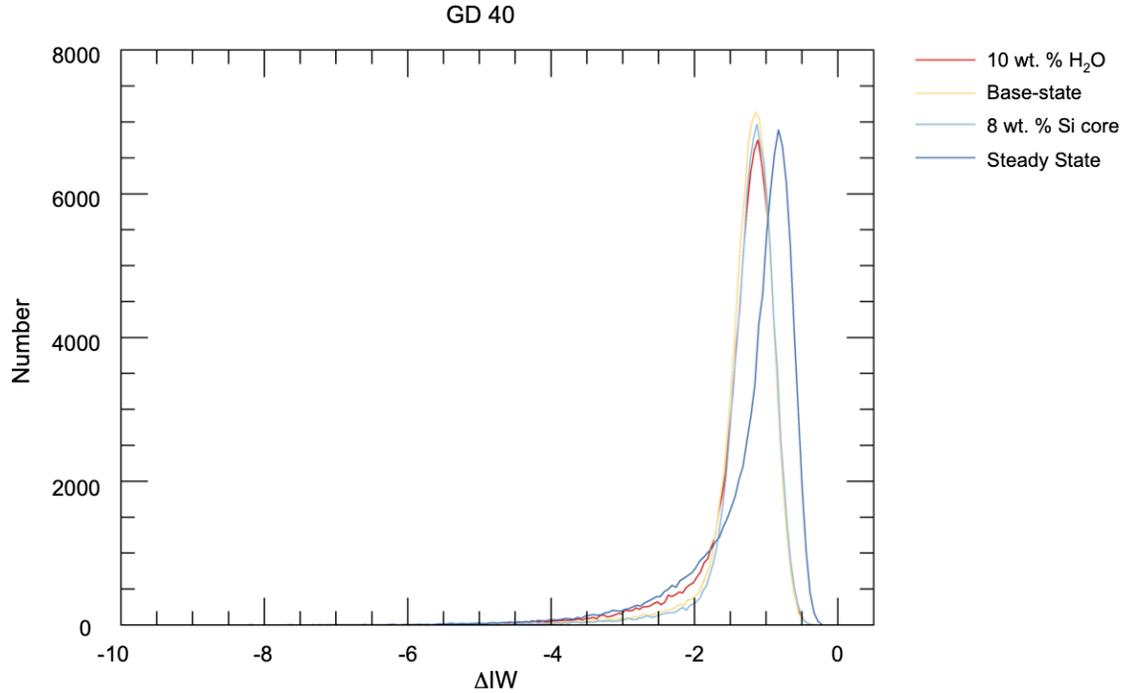

**Fig. S5. Variations in calculated oxygen fugacities for WD GD 40.** Changes in $\Delta$IW are shown for different calculations for WD GD 40, described above. Our base-state calculation is shown in yellow. A calculation using steady-state elemental abundances is shown in blue, a calculation with the rocky body accreting onto GD 40 having 10 wt. % water is shown in red, and a calculation with 18% of the Si in the rocky body allocated to the core (in Earth, this amount corresponds to ~ 8 wt. % Si in the core) is shown in light blue.

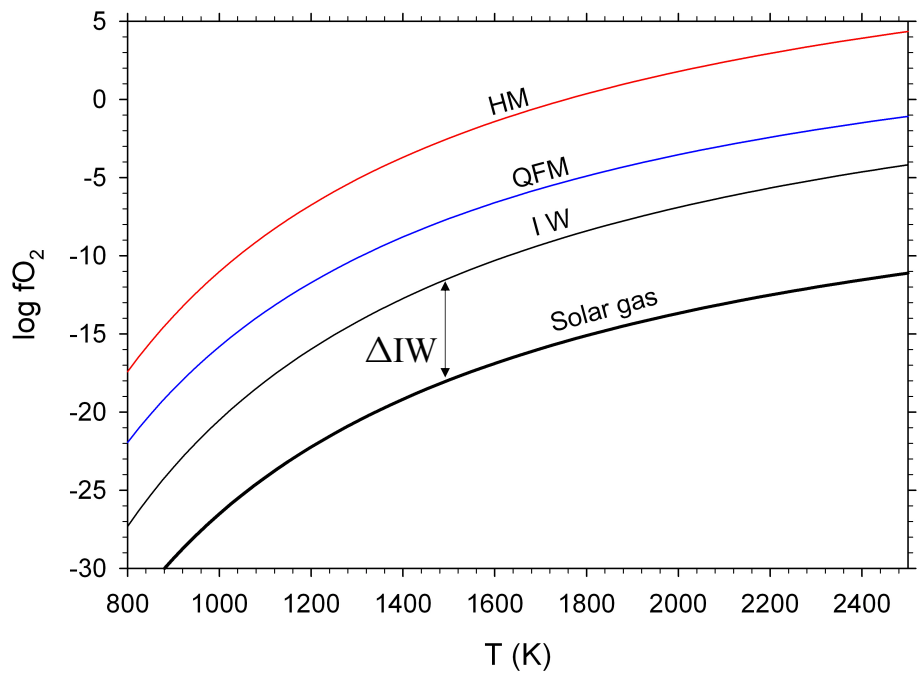

**Fig. S6. Logarithm of oxygen fugacity vs. temperature for relevant reactions.** The $\log(f_{O_2})$ versus temperature curves are those for iron-wüstite (IW), representing the equilibrium reaction $Fe + \frac{1}{2}O_2 = FeO$, quartz-fayalite-magnetite (QFM), representing the reaction $3Fe_2SiO_4 + O_2 = 2Fe_3O_4 + 3SiO_2$, and hematite-magnetite (HM), representing the reaction $2Fe_3O_4 + \frac{1}{2}O_2 = 3Fe_2O_3$ (*70*). These curves are compared with the analogous curve for a gas of solar composition (*71*). $\Delta$IW for the solar gas is indicated.

**Table S1 | Data for white dwarfs in this study.** WD types are as described in Methods and the order of the letters is based on which constituent is most dominant in the atmosphere.

| White Dwarf ICRS coordinates, J2000 | Type | Temperature (K) | ΔIW | References |
|---|---|---|---|---|
| GD 40 03 02 53.10 -01 08 33.80 | DBZA | 15,300 | $-1.17^{+0.25}_{-0.33}$ | (8, 12) |
| GD 40 (steady state) | | | $-0.95^{+0.29}_{-0.72}$ | (8, 12) |
| WD 1536+520 15 37 25.73 +51 51 26.8 | DBAZ | 20,800 | $-1.37^{+0.28}_{-0.32}$ | (11) |
| SDSS J0738+1835 07 38 42.57 +18 35 09.7 | DBZA | 13,950 | $-1.14^{+0.17}_{-0.18}$ | (13) |
| WD 1145+017 11 48 33.63 +01 28 59.4 | DBZ | 16,900 | $-0.59^{+0.13}_{-0.16}$ | (14) |
| SDSS J1043+0855 10 43 41.53 +08 55 58.3 | DAZ | 18,330 | $< -1.21$ | (10) |
| WD 1226+110 12 28 59.93 +10 40 33.0 | DAZ | 20,900 | $-0.63^{+0.31}_{-0.43}$ | (9) |

**Data S1 | Input data and results for this study.** This file contains information for the WDs and Solar System bodies used in this study, including elemental abundances used in the bootstrap and analytical methods (Data S2 and S3). The file contains resulting oxygen fugacities derived using both the bootstrap and analytical methods described in Materials and Methods. Also, this file contains input concentrations used to make Fig. 1.

**Data S2 | IDL data analysis code.** This IDL .pro file contains the code to calculate oxygen fugacities using the bootstrap method described in Materials and Methods.

**Data S3 | Fortran data analysis code.** This Fortran source code was used to calculate oxygen fugacities using the analytical method described in Materials and Methods.